\documentclass[conference]{IEEEtran}
\pdfoutput=1
\IEEEoverridecommandlockouts

\usepackage{tikz-network}

\usepackage{amsmath}
\usepackage{graphicx}

\usepackage[colorlinks=true,allcolors=blue]{hyperref}
\usepackage{amsthm}

\usepackage{lipsum}
\usepackage{flushend}

\usepackage{todonotes}

\theoremstyle{definition}

\theoremstyle{definition}

\usepackage{xspace}

\title{Wild SBOMs: a Large-scale Dataset of Software Bills of Materials from Public Code}
\author{
  \IEEEauthorblockN{Luís Soeiro\IEEEauthorrefmark{1}, Thomas Robert\IEEEauthorrefmark{1}, Stefano Zacchiroli\IEEEauthorrefmark{1}}
  \IEEEauthorblockA{\IEEEauthorrefmark{1}LTCI, Télécom Paris, Institut Polytechnique
		 de Paris, France\\
    \{luis.soeiro, thomas.robert, stefano.zacchiroli\}@telecom-paris.fr}
  \thanks{Supported by the industrial chair Cybersecurity for Critical Networked Infrastructures (cyberCNI.fr) with support of the FEDER development fund of the Brittany region, France.
    This work was made possible by Software Heritage, the great library of source code: \url{https://www.softwareheritage.org}.
    Special thanks to Valentin Lorentz for his help with data extraction.}
}

\begin{document}
\maketitle

\begin{abstract}
Developers gain productivity by reusing readily available Free and Open
	Source Software (FOSS) components. Such practices also bring some
	difficulties, such as managing licensing, components and related
	security. One approach to handle those difficulties is to use Software
	Bill of Materials (SBOMs). While there have been studies on the
	readiness of practitioners to embrace SBOMs and on the SBOM tools
	ecosystem, a large scale study on SBOM practices based on SBOM files
	produced in the wild is still lacking. A starting point for such a
	study is a large dataset of SBOM files found in the wild. We introduce
	such a dataset, consisting of over 78 thousand unique SBOM files,
	deduplicated from those found in over 94 million repositories. We
	include metadata that contains the standard and format used, quality
	score generated by the tool \textit{sbomqs}, number of revisions,
	filenames and provenance information. Finally, we give suggestions and
	examples of research that could bring new insights on assessing and
	improving SBOM real practices.

\end{abstract}


\section{Introduction}
\label{sec:intro}

Modern software development reuses relevant amounts of code from third
parties, mainly in the form of Free and Open Source (FOSS) code, to
build applications~\cite{Buchkova2022}. FOSS ecosystems can be very
large, from thousands of components to tens of thousands when
accounting for their dependencies~\cite{Decan2018}. While gains from
reusing FOSS code can reach up to trillions of
dollars~\cite{Hoffmann2024}, there are also some difficulties to be
handled, such as licensing requirements~\cite{Phipps2020}, repository
and component governance~\cite{Harutyunyan2019a}, and security
considerations~\cite{Lin2023}. One of the mechanisms that have been
proposed to aid with those issues and to provide more transparency to
software projects is the use of Software Bill of Materials
(SBOM)~\cite{Stalnaker2024}, an inventory of all third-party components
and dependencies used in an application.~\cite{Mirakhorli2024}.

Since the publication of the Software Identification Tags (SWID Tags)
in 2009~\cite{Waltermire2016}, the Software Package Data Exchange
(SPDX) SBOM introduction in 2010~\cite{r_spdxabout2024}, and the
CycloneDX SBOM standard prototype in 2017~\cite{r_cdxhistory2024},
SBOMs have gained traction. In the wake of major software supply chain
security incidents (e.g., SolarWinds, Log4J), the US National
Telecommunications and Information Administration (NTIA), following the
U.S. Presidential Order 14028, started to promote the use of
SBOMs~\cite{Stalnaker2024}, and it has become a requirement for
supplying software to the U.S. government~\cite{Zahan2023}. In the
E.U., the Cyber Resilience Act (CRA) proposal of 2022 also brings a
similar requirement~\cite{DallaPreda2024}.

There has been studies on the benefits and challenges of SBOM
adoption~\cite{Zahan2023,Otoda2024}, on the practitioners' views on the
subject~\cite{Stalnaker2024}, and on the difficulties of generating
correct SBOM files using SBOM creation tools~\cite{Balliu2023}.
According to a Linux Foundation's survey from 2022, 20\% of the
organizations interviewed are already producing SBOMs and 40\% are
consuming them in production~\cite{r_lf2022}. Research on real SBOM
files found in the wild can help to evaluate the state of those
practices. Existing SBOM datasets mined from repositories contain a
maximum of 1,151 files, but lack diversity (see \ref{sec:rw}). Easily
accessible, more diverse datasets, with more samples are
needed~\cite{TorresArias2023}. Analysis of such a corpus can provide
insights to real world SBOM usage aspects, including standards
adherence and overall quality. Additionally, it can facilitate the SBOM
tools ecosystem, by providing developers with material to test,
evaluate and benchmark tools.

\paragraph*{Contributions and use cases} We introduce a large and diverse
dataset of SBOM files that came from public version control systems
(VCS). The package is comprised of two parts:

\begin{enumerate}
	\item A deduplicated dataset of 78,612 unique SBOM files, that were found on
	      94,618,356 unique source repositories, distributed in 1,782 unique
	      forges and package repositories (\textit{forges} for short);
	\item Mined metadata, including the SBOM standard adopted (e.g., SPDX,
	      CycloneDX), the file format (e.g., xml, json, tag-value, yaml), a
	      quality score generated by the \textit{sbomqs}
	      tool~\cite{r_sbomqs2024}, provenance information for each SBOM file,
	      and all the different filenames each one had being observed with.
\end{enumerate}

The dataset can be used to support use cases like: (a) large scale
analysis of SBOM adoption, most used standards, and file formats,
possibly segmented by originating forge, creator tool and other
properties; (b) analysis of SBOM quality in the wild; (c) benchmarking
of the SBOM tools, by evaluating their effectiveness and correctness of
their functionalities, such as SBOM consumption, conversion, and
validation; (d) software composition analysis; (e) vulnerability
analysis.

\paragraph*{Data availability} The dataset is released as open data,
and the related code as free and open source code. The replication
package allows the dataset to be recreated from scratch. It is
available from Zenodo~\cite{r_zenodo},
as a tar archive containing a
directory tree with all the SBOM files, a set of CSV files containing
the metadata, the software used to generated it, and detailed usage and
reproducibility instructions (\textit{see README.md} file).

\section{Methods and Reproducibility}
\label{sec:methods}

\begin{figure}[h]
	\centering
	\includegraphics[width=0.5\textwidth]{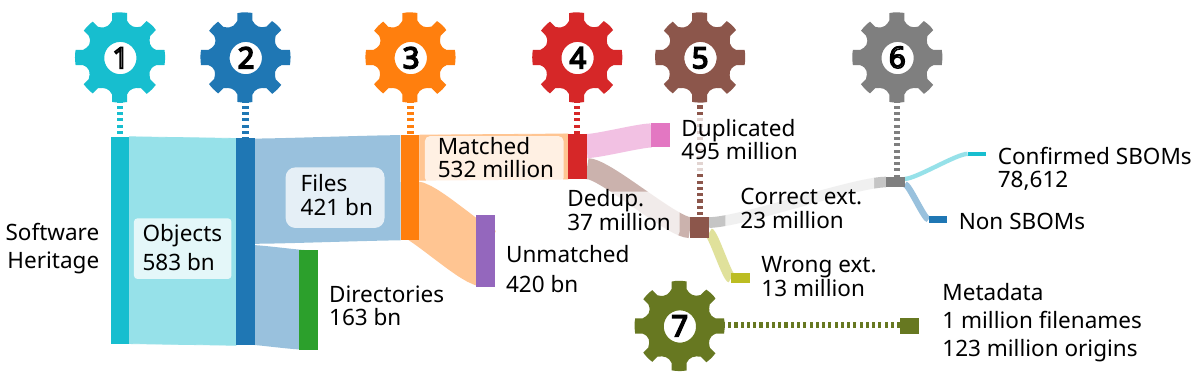}
	\caption{Overview of the SBOM dataset creation}
	\label{fig:dataset-creation}
\end{figure}

We have adopted the following criteria for building the dataset: (a)
diversity (i.e., it should include as many different forges and package
repositories as possible); (b) availability (i.e., it should be
publicly available); (c) FOSS based (i.e., reflecting FOSS software
development and artifacts); (d) history (i.e., the SBOM file should be
part of the repository, statically embedded on its commit history).
Consequentially, we have opted to search the Software Heritage
Archive~\cite{DiCosmo2017} (SWH). If a SBOM file is found in the public
VCS of any of the main software forges, there is a high probability of
being found in SWH, which archives over 4 billion commits and over 324
million diverse public software repositories, at the date of this
writing~\cite{r_swh2024}. Since there is no clear indication of which
artifacts are SBOM files, we have defined a strategy to mine the
archive, acquire possible candidates and then filter out non-SBOM
files. The process is depicted in Figure \ref{fig:dataset-creation} and
each of the steps identified are detailed below.

\subsubsection{Download of indexes} First we download the \textit{Optimized Row Columnar} (ORC) files that
correspond to directory entries and the mapping of the intrinsic
identifiers we will use (i.e., SHA1, and SWHID) for the release version
"2024-08-23" of the Software Heritage graph dataset~\cite{Pietri2019}
on Amazon S3. The data contains 583,607,093,407 files and directory
entries.

\subsubsection{Filter for filenames} We then load those ORC files into Apache Spark~\cite{Salloum2016} to
retain only file entries. This results in 420,647,294,867 file
references.

\subsubsection{Search for regular expressions}
The goal is to obtain as much candidates for SBOM files as possible.
However, at such scale it is necessary to reduce the set of possible
values. We use Spark to look for file names typically used to store
SBOMs. Accordingly, we search any file name that contains, in a case
insensitive way, at least one of the substrings: \textit{spdx},
\textit{swid}, \textit{bom}, \textit{cyclone}, \textit{cdx}, or
\textit{dx}. This execution results in 531,729,785 items.

\subsubsection{Deduplication of findings}
\label{sub:deduplication}
The same SBOM content may be observed in different \textit{commits} of
the same repository, in different repositories, or with different file
names. We save the list of file names to add to the metadata later (see
section \ref{sub:metadata}). Then we drop duplicate contents (i.e.,
duplicate SHA1 hash values). This results in 37,036,596 candidates.

\subsubsection{Removal of records with non-SBOM file extensions}
We filter out records that contain file extensions which are known to
belong to other than SBOM file formats (e.g.,
\textit{.dll},\textit{.so} for dynamic libraries; \textit{.odt}, for
LibreOffice text documents; \textit{.md} for markup documents). We
obtain a list of 991 such extensions from the Apache mime-types
file~\cite{r_apachemime}. Then we obtain 23,590,858 records.

\subsubsection{Download and evaluation of candidate SBOM files}
We use the SHA1 hash values present in the candidate records to
download the candidate files from Amazon S3, and them validate that
their contents are not corrupted. Then we evaluate them to make sure to
retain only real SBOM files. For this evaluation we have selected the
\textit{SBOM quality score (sbomqs)} tool~\cite{r_sbomqs2024}, since it
has already been used to score over 28 thousand SBOM files on the
\textit{sbombenchmark.dev} site~\cite{r_sbombench2024}. We define that
valid SBOM files for inclusion in the dataset are those that don't
receive a \textit{failed} result by the SBOM scoring tool. For those
SBOMs included in the dataset, we store the results of the tool as
metadata: quality score, SBOM standard, SBOM format and SBOM version.
After the filtering process we obtain 78,612 confirmed SBOM files.

\subsubsection{Metadata addition}
\label{sub:metadata}
We then select only those file names obtained in the deduplication step
(section \ref{sub:deduplication}) whose SHA1 hash values match the SHA1
hash values from the valid SBOM files. This results in 1,168,328
records with file names. We also traverse the SWH Graph to find all the
repositories (i.e., \textit{origins} in SWH) that have been observed
for each SBOM file. The resulting data has 123,826,651 records with
origins. Finally, we query SWH to obtain the earliest observed commit
(i.e., \textit{revision} in SWH) where the SBOM was observed, the
earliest observed timestamp, and the total number of observed commits
for each SBOM file. We sum all commits and observe that the SBOM files
are distributed among 2,232,518,895 commits.

\section{Description of the dataset}
\label{sec:data-model}
The dataset contains a collection of SBOM files and the associated
metadata, totalling 14 GB. We compress all elements of the dataset with
\textit{Zstandard}~\cite{rfc8878} to reduce the storage space, bringing
it down to 5.6 GB.

\paragraph*{SBOM files} The file \textit{sbom-files.tar.zstd} contains all the
78,612 SBOM files. When extracted they occupy 12 GB of storage space. They are
organized under a 1-level deep directory structure based on the first
character of their SHA1 hash value, e.g.,
\text{sbom-files/a/a01269419c76765dfa79499597e8eac79787450d}. The files
are deduplicated based on the SHA1 value (i.e., even if the same file
was observed in different origins and different revisions, it will
appear only once in the dataset).

\begin{figure}[h]
	\centering
	\includegraphics[width=0.5\textwidth]{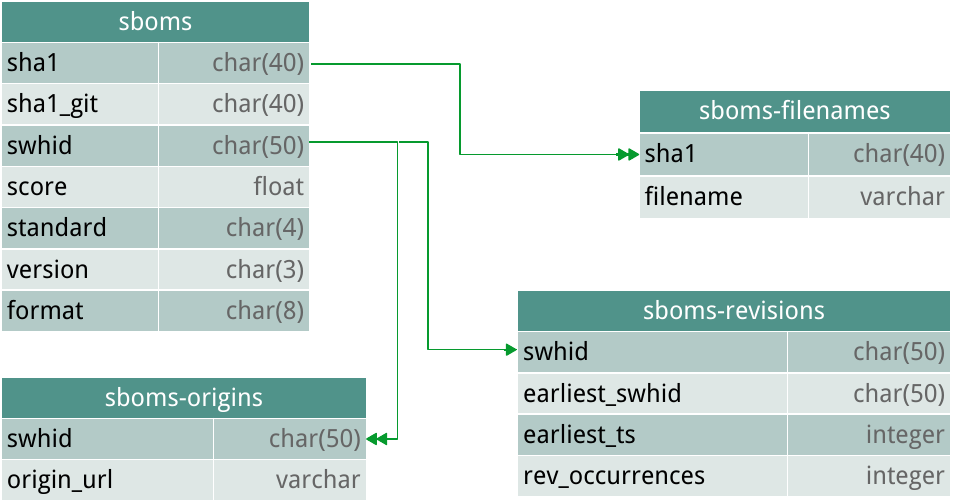}
	\caption{Relational model for the SBOM dataset metadata}
	\label{fig:datamodel}
\end{figure}

\paragraph*{Metadata}
SBOM Metadata are provided as a set of 4 CSV~\cite{rfc4180} textual
files, which correspond to 4 relational database tables, shown in
Figure \ref{fig:datamodel}. The tables are described below, with the
indication of the corresponding file:

\textbf{sboms} (\textit{sboms-01.csv.zst}) is the main table of the
dataset and describes the main characteristics of each SBOM file. The
\textit{sha1} field is the SHA1 hash value of the content of the SBOM
file, and its filename in the dataset; \textit{sha1\_git} is the
original Git commit value where the SBOM was found; \textit{swhid} is
the file's Software Heritage persistent identifier
(SWHID)~\cite{DiCosmo2018}; \textit{score} is the evaluation score
given by \textit{sbomqs} and ranges from $0$ to $10$; \textit{standard}
indicates the SBOM standard that is used for the SBOM: \textit{spdx}
(i.e., the SPDX stanard) or \textit{cdx} (i.e. the CycloneDX standard);
\textit{format} indicates the layout format of the SBOM: \textit{json},
\textit{xml}, or \textit{tag-value}.

\textbf{sboms-filenames} (\textit{sboms-02-filenames.csv.zst}) provides
the repository maintainer's choice for naming the SBOM, i.e., the file
name that have been observed for each SBOM file. \textit{sha1} is the
SHA1 hash value for the SBOM contents and a foreign key to the
\textit{sboms} table; \textit{filename} is an observed filename for
that SBOM file, e.g., \textit{camel-sbom.xml}, \textit{bom.xml}. There
can be many filenames observed for each SBOM content, i.e., many rows.
A SBOM file which is renamed and committed to a VCS multiple times
without content change will have one row for each distinct file name.

\textbf{sboms-origins} (\textit{sboms-03-origins.csv.zst}) shows all
the repositories where each SBOM file has been observed in. \textit{swhid} is
the file's SWHID and a foreign key to the \textit{sboms} table;
\textit{oring\_url} denotes the repository where the SBOM file was observed
in the past, for example, \textit{https://codeberg.org/cap\_jmk/tinkabell.git}.
There can be many repositories for each SBOM file, e.g. repository
\textit{forks}.

\textbf{sboms-revisions} (\textit{sboms-04-revisions.csv.zst}) provides
historical information. \textit{earliest\_swhid} gives the SWHID of the
oldest known public commit that contained the SBOM file;
\textit{earliest\_ts} is the commit timestamp as Unix time;
\textit{rev\_occurrences} shows the total number of commits that
contain the SBOM file, as known by SWH. For example,
swh:1:cnt:0004b472fcd003bcbd29544a534d1f24afd0f3f2 denotes an instance
of a SBOM for which the oldest commit is from May 8, 2022, and has a
history of 52 revisions. By looking up the SHA1 value in \textit{sboms}
and consulting the file in \textit{sbom-files} we can see it is a
CycloneDX 1.3 JSON SBOM file for a Go language application.

\section{Dataset Use Cases}
This dataset can be used to study SBOM creation practices in the wild,
e.g., when they were created, main standards and formats, tools used,
contents that are present. It can also be used to evaluate and
benchmark SBOM related tools. Other possibilities include vulnerability
analysis at large by searching the SBOM contents on vulnerability
databases. We give some examples related to understanding SBOM
practices below. Details for each of the examples are found in the
\textit{Jupyter Notebooks} provided with the dataset.

\subsection{Standards}
\label{sec:standards}

Since NTIA, who is pushing for SBOM adoption in the industry, has not
endorsed any specific standards~\cite{Stalnaker2024}, it is up to the
developers to decide what standard to adhere to. SWID Tags was proposed
first and is now the standard ISO/IEC 19770-2:2015~\cite{r_ntiaswid}.
SPDX is a project by the Linux Foundation and was standardized as
ISO/IEC 5962:2021~\cite{r_spdxabout2024}. CycloneDX is younger than
both~\cite{r_cdxhistory2024}. Do those elements affect general
adoption? One first step to answering that is to verify what standards
and file formats are really being used by developers.

\begin{figure}[h]
	\centering
	\includegraphics[width=0.5\textwidth]{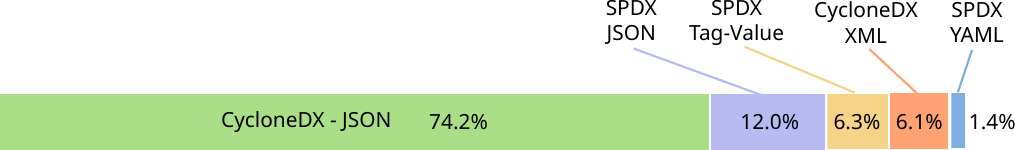}
	\caption{Distribution of SBOM standards and file formats}
	\label{fig:standards}
\end{figure}

We use the Python Pandas library~\cite{mckinney2011} to load the
\textit{sboms} table and count the frequency of standards and formats
found in the wild. Figure \ref{fig:standards} show the results. This is
a first approximation of the analysis because we are studying the
\textit{deduplicated} set only. Although a more comprehensive study
might give some weights to SBOMs that have more origins and revisions,
this first version considers the underlining choices of the original
creators. It seems to be surprising that CycloneDX is the most used
standard, while being the youngest proposal. Furthermore we see that
there are no SWID Tags based SBOMs at all.

\subsection{Filenames}
We haven't found in the SPDX specification guidelines for file
extension or file naming conventions. OpenSSF Best Practices document
recommends to add the extension \textit{*.spdx} for tag-value format or
\textit{*.spdx.\{json, xml, yml, yaml, rdf, rdf-xml\}
}~\cite{r_openssf}. CycloneDX recommends the adoption of the filenames
\textit{bom.\{json, xml\}} or the extensions \textit{*.cdx.\{json,
	xml\}}~\cite{r_cdxoverview}. We will see what practitioners adhere to.

\begin{figure}[h]
	\centering
	\includegraphics[width=0.5\textwidth]{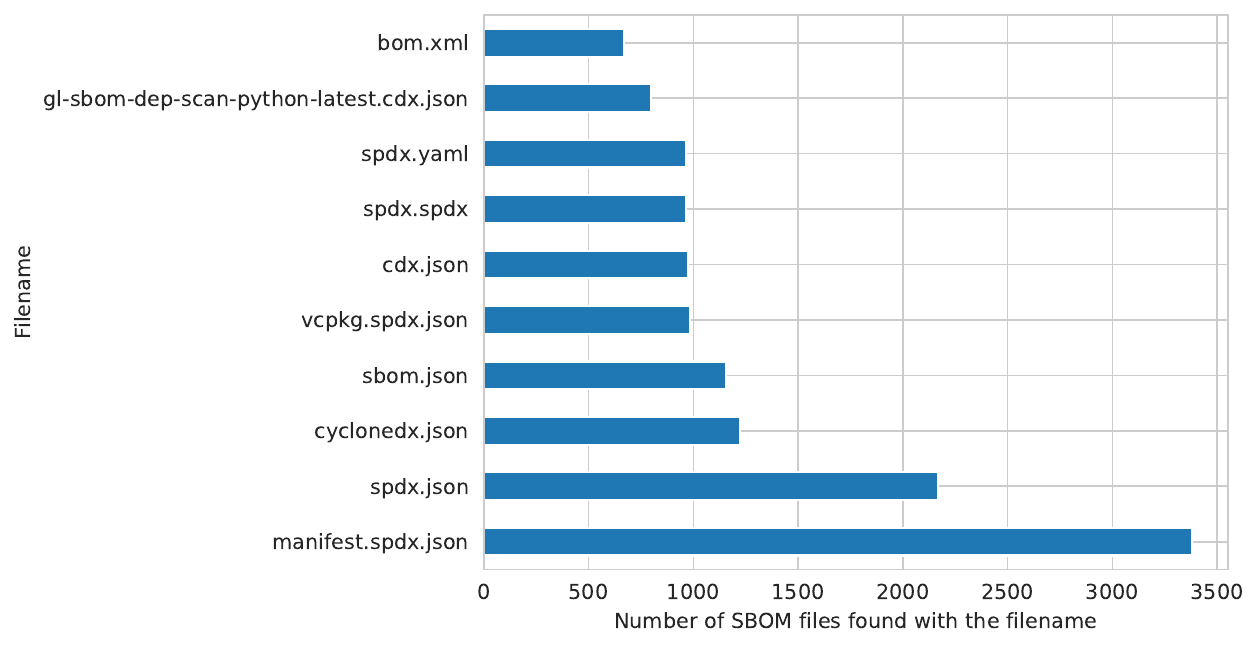}
	\caption{The most popular filenames for SBOM files}
	\label{fig:popular-filenames}
\end{figure}

We use Pandas to load the table \textit{sboms-filenames}, count the
number of occurrences and display the 10 most used file names. Figure
\ref{fig:popular-filenames}. For this example, we look at all filenames
that are used in all repositories throughout all commits.

\subsection{Popular forges and package repositories}
The dataset contains 94,618,356 unique origins. Each of those origins
point to one of 1,782 unique forges. We examine how SBOM practitioners'
works are distributed along the forges.

\begin{figure}[h]
	\centering
	\includegraphics[width=0.5\textwidth]{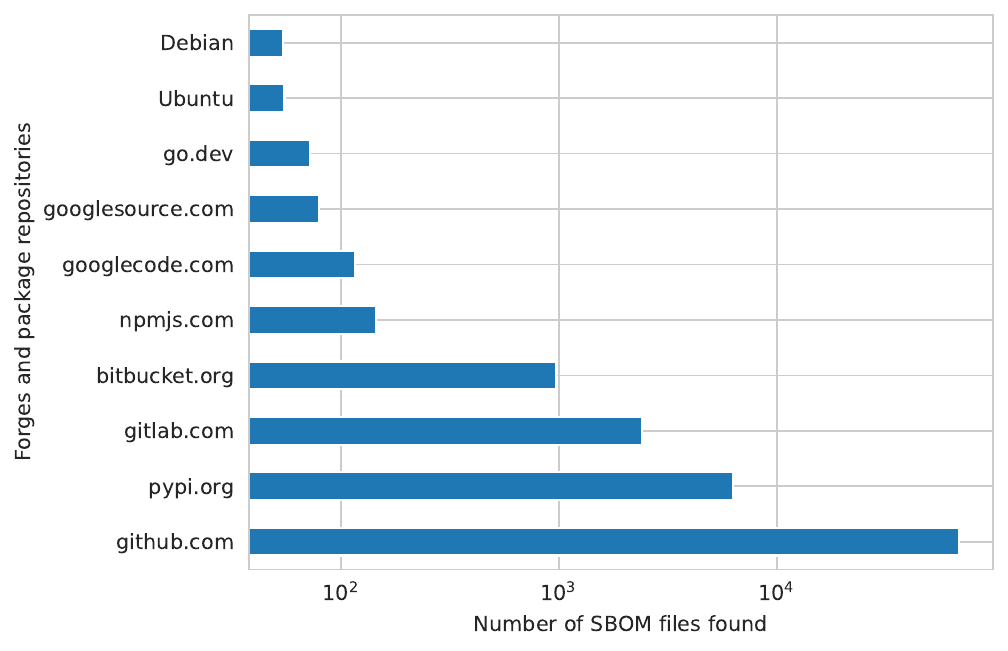}
	\caption{Forges with the most SBOM files}
	\label{fig:popular-forges}
\end{figure}

We use Pandas to load the table \textit{sboms-origins} and remove the
duplicates of \textit{origin\_url}. Then we use the library
\textit{tldextract}~\cite{r_tldextract} to help extract the part that
correspond to the forge. In total, the practitioners have placed SBOMs
in 1,783 unique forges. The top ten forges that contributed with the
most SBOM files to the dataset are shown in figure
\ref{fig:popular-forges}.

\section{Limitations}

We took the approach of searching for SBOM files that are stored in
public VCS. However, most of the forges also have a way (e.g., Github
assets) for the contributors to release software artifacts that are
stored outside the version control system and SBOM files could be
placed there as well. Additionally, the popular forges
Github~\cite{r_githubsbom2024} and Gitlab~\cite{r_gitlabsbom2024} have
recently announced services to integrate dynamic SBOM generation in
their pipelines. Our approach in this work doesn't account for those
SBOM files, as they are released outside of VCS repositories. There is
still no consensus to where SBOM files should be stored, and it is
possible to argue for storing them on either place. If practitioners
favor the approach that requires the least amount of work, they may use
the platform's SBOM creation functionalities, and fewer SBOM files will
be found in public VCS repositories. However, storing SBOM files inside
the VCS repositories makes them independent of that specific forge
(i.e., cloning the VCS will retrieve all resources). Furthermore, we
sampled some repositories from the dataset and looked at their VCS
contents in SWH. There are contributors that include the SBOM file in
the same directory of the release notes, inside the VCS, where they are
kept as part of the history along with the source files.

\section{Related Work}
\label{sec:rw}

Previous works have created SBOM datasets from existing projects, as
part of a broader research to evaluate SBOM generating tools for
feature comparison~\cite{Dalia2024}, on Docker
images~\cite{Kawaguchi2024}, for security
purposes~\cite{Kagizmandere2024}, or for accuracy
assessment~\cite{Balliu2023,Mirakhorli2024,Yu2024,Rabbi2024}. There are
two main differences that our work presents. First, our work doesn't
generate the SBOM files. We aggregate those committed by practitioners
to VCS repositories, which possibly provides for more diversity in the
conditions of generation (e.g., time of creation, origin, authors,
generation tools) to better be able to assess SBOM practices; Second,
while the number of SBOM files in those works range from 2 to 7,876,
our dataset is much larger by an order of magnitude.

There are also related works that gather SBOM files in the wild. Ambala
has mined SBOM files from 18 software repositories for security
analysis~\cite{Ambala2024}. The dataset "bom-shelter" contains over 50
SBOM files found by using a web site specialized on source code
search~\cite{r_chain2022}. Nocera et al have selected SBOM creation
tools that have Github repositories and then they have used Github's
API to locate software repositories that depend on those tools, finding
186 SBOM files in the wild~\cite{Nocera2023}. O’Donoghue has obtained
1,151 SBOM files from Interlynk's SBOM
sbombenchmark.dev~\cite{r_sbombench2024}, by accessing directly its
Amazon S3 repository, to check them for vulnerabilities using SBOM
tools~\cite{Odonogue2024}. Our approach differs from these works on the
diversity and quantity of SBOM files found in the wild. While most
studies rely on sourcing one or a few well known software forges and
collections (e.g., GitHub, sbombenchmark.dev), our work represents
1,782 unique forges and 94,618,356 unique source repositories,
providing a larger and a much more diverse dataset.

\section{Conclusion}
\label{sec:conclusion}

We have introduced a dataset of 78,612 unique SBOM files found in the
wild. They were observed in 94,618,356 repositories, spread in 1,782
unique forges. We also included the metadata: \textit{sha1},
\textit{sha1\_git}, \textit{swhid}, SBOM score, SBOM standard, SBOM
version, SBOM format, and provenance information. We've given some
suggestions of future usage and we have presented some concrete
examples.

\paragraph*{Future work} The goal is to enlarge the dataset and to
improve the available
metadata. We will include newer SBOMs found in the wild and also add
metadata related to quality criteria as measured by diverse SBOM tools.

\end{document}